# THE RELATIONSHIP AMONG THE STANDARD DEVIATION OF A DATASET AND DE STANDARD DEVIATION AND AVERAGE TWO PART OF THIS SET.

JOSE FAUSTO DE MORAIS, PhD

*Faculty of Mathematics of University of Uberlândia, MG, Brazil*

**ABSTRACT**

Meta-analysis involves combining summary information for related but independent studies. It uses different relationship to combine position measure as well as dispersion measures. The objective of this study is to discuss a relationship among the standard deviation of a data set and the standard deviation and mean of two part of this set. The problem was proposed in a systematic review with meta-analysis that combined two studies with missing data.

## INTRODUCTION

Meta-analysis may be defined as the quantitative review and synthesis of the results of related but independent studies[1]. The objective of a meta-analysis can be several-fold. By combining information over different studies, an integrated analysis will have more statistical power to detect a treatment effect than an analysis based on only one study.

As in primary research, a meta-analysis begins with a well-formulated question and design. (1) What are the study objectives? Is the objective of the study to validate results in a broader population? (2)What are the operational definitions of the research outcome, the treatment or intervention, and the population? (3)What types of designs will be included in the search? Will only randomized trials testing the research hypothesis be included or will be results from non-experimental studies be permitted? Will randomized trials with poor compliance be included?

The answer to these questions impact on the methods of review, the modes of statistical inference, and the interpretation of the results. Once the primary studies have been collected and coded, the meta-analyst needs to identify a summary measure common to all studies and subsequently combine the measure.

In several situations the analyst wants to combine two or more studies and calculate the standard deviation of the combined group when he knows only the average, the standard deviation and the size of each studies. The objective this paper is present a formula for to combine the standard deviation of two studies.

**FORMULATING**

Consider the set

$$X = \{x_1, x_2, \ldots, x_R\} \text{ and } Y = \{y_1, y_2, \ldots, y_A\}.$$

The statistical literature[2,3,4] show us that

$$S_X^2 = \frac{1}{R-1}\left[\sum_1^R x_i^2 - R\bar{x}^2\right] \quad \text{and} \quad S_Y^2 = \frac{1}{A-1}\left[\sum_1^A y_i^2 - A\bar{x}^2\right] \quad (1).$$

Where $S_X^2$ and $S_Y^2$ are sample standard deviation of X and Y respectively. If we define the set with repetition of elements XvY, where

$$XvY = \{x_1, x_2, \ldots, x_R, y_1, y_2, \ldots, y_A\}$$

we want to obtain a expression for $S_{XvY}^2$ in function of

$$\bar{x}, \quad \bar{y}, \quad S_X^2, \quad S_Y^2, \quad R \quad \text{and} \quad A$$

It's possible to show that

$$S_{XvY}^2 = \frac{1}{R+A-1}\left[(R-1)S_X^2 + (A-1)S_Y^2 + R\bar{x}^2 + A\bar{y}^2 - \frac{(R\bar{x}+A\bar{y})^2}{R+A}\right]$$

Where $\bar{x}$ and $\bar{y}$ are the average, $S_X^2$ and $S_Y^2$ are the standard deviation, and R and A are the size of the X and Y set, respectively. $S_{XvY}^2$ are standard deviation of the XvY set

**DEDUCTION**

If we isolate the sums in (1), we will have

$$\sum_1^R x_i^2 = (R-1)S_X^2 + R\bar{x}^2 \quad \text{and} \quad \sum_1^A y_i^2 = (A-1)S_Y^2 + A\bar{y}^2 \quad (2)$$

Expanding the expression $\sum_1^{R+A}(XvY)_i^2$ we will have

$$\sum_1^{R+A}(XvY)_i^2 = x_1^2 + x_2^2 + \ldots + x_R^2 + y_1^2 + y_2^2 + \ldots + y_A^2 = \sum_1^R X_i^2 + \sum_1^A Y_i^2 \quad (3)$$

Putting (2) in (3) we will have

$$\sum_1^{R+A}(XvY)_i^2 = (R-1)S_X^2 + R\bar{x}^2 + (A-1)S_Y^2 + A\bar{y}^2 \quad (4)$$

Again, the statistical literature permits define

$$S^2_{XvY} = \frac{1}{R+A-1}\left[\sum_{1}^{R+A}(XvY)^2_i - (R+A)\overline{(XvY)}^2\right] \quad (5)$$

but

$$\overline{XvY} = \frac{\sum_{1}^{R+A}(XvY)_i}{R+A} = \frac{x_1+x_2+...+x_R+y_1+y_2+...+y_A}{R+A} = \frac{\sum_{1}^{R}x_i + \sum_{1}^{A}y_i}{R+A} = \frac{R\overline{X}+A\overline{Y}}{R+A} \quad (6)$$

Putting (4) and (6) in (5) and simplifying the expression, we will obtain

$$S^2_{XvY} = \frac{1}{R+A-1}\left[(R-1)S^2_X + (A-1)S^2_Y + R\overline{x}^2 + A\overline{y}^2 - \frac{(R\overline{x}+A\overline{y})^2}{R+A}\right]$$

**COMMENTARY**

in the context of meta-analysis, if X is a set of values in study and Y is the set of values lost in the study, the formula presented allows relate the standard deviation and average of X and Y with standard deviation of XvY. We are working in generalization of formula, but this is a discussion for complete paper.

---